\documentclass[final,3p,times,sort&compress]{elsarticle}

\usepackage{multirow,setspace,amssymb,amsmath,graphicx,color,rotating,subfigure,url,lineno,natbib}
\usepackage{textcomp}

\journal{Physics Letters A} 

\begin{document}

\begin{frontmatter}

\title{Modeling interaction of trading volume in financial dynamics}%
\author[BS,ZIMP,RCE]{F. Ren}
\author[ZIMP]{B. Zheng \corref{cor}}
\cortext[cor]{Corresponding author. Address: Physics Department,
Zhejiang University, Hangzhou 310027, China.}
\ead{zheng@zimp.zju.edu.cn} %
\author[ZIMP]{P. Chen}

\address[BS]{School of Business, East China University of Science and Technology, Shanghai 200237, China}
\address[ZIMP]{Zhejiang Institute of Modern Physics, Zhejiang University, Hangzhou 310027, China.}
\address[RCE]{Research Center for Econophysics, East China University of Science and Technology, Shanghai 200237, China}

\begin{abstract}
A dynamic herding model with interactions of trading volumes is
introduced. At time $t$, an agent trades with a probability, which
depends on the ratio of the total trading volume at time $t-1$ to
its own trading volume at its last trade. The price return is
determined by the volume imbalance and number of trades. The model
successfully reproduces the power-law distributions of the trading
volume, number of trades and price return, and their relations.
Moreover, the generated time series are long-range correlated. We
demonstrate that the results are rather robust, and do not depend on
the particular form of the trading probability.
\end{abstract}

\begin{keyword}
 Econophysics; Scaling laws; Complex systems
 \PACS{89.65.Gh, 89.75.Da, 89.75.-k}
\end{keyword}

\end{frontmatter}

\section{Introduction}
\label{S1:Intro}

In statistical analysis of financial markets, much attention has
been drawn to the study of the stock prices
\cite{Mandelbrot-1963-JB,Fama-1965-JB,Mantegna-Stanley-1995-Nature,Lux-1996-AFE,Ghashghaie-Breymann-Peinke-Talkner-Dodge-1996-Nature,Plerou-Gopikrishnan-Amaral-Meyer-Stanley-1999-PRE,Plerou-Gopikrishnan-Amaral-Gabaix-Stanley-2000-PRE,Gabaix-Gopikrishnan-Plerou-Stanley-2003-Nature,Giardina-Bouchaud-Mezard-2001-PA,Liu-Gopikrishnan-Cizeau-Meyer-Peng-Stanley-1999-PRE,Mantegna-Stanley-2000,Bouchaud-Potters-2000,Sornette-2003}.
Denoting $p(t)$ as the price of a given stock or financial index,
the price return $r(t)$ is defined as the change of the logarithmic
price in a time interval $\Delta t$, i.e., $ r(t)\equiv \ln p(t)-\ln
p(t-\Delta t)$. A power-law tail with an exponent $\xi_r\cong3.0$ is
found in the cumulative probability distribution of price returns
\cite{Plerou-Gopikrishnan-Amaral-Meyer-Stanley-1999-PRE,Plerou-Gopikrishnan-Amaral-Gabaix-Stanley-2000-PRE,Gabaix-Gopikrishnan-Plerou-Stanley-2003-Nature},
which indicates that the large price fluctuation is more common than
one might naively expect. Another important statistical property
basically observed in financial markets is the long-range
correlation of the volatility which is simply defined as the
magnitude of the price return
\cite{Giardina-Bouchaud-Mezard-2001-PA,Liu-Gopikrishnan-Cizeau-Meyer-Peng-Stanley-1999-PRE}.
Many efforts have been devoted to the understanding of the financial
markets along this direction, with both phenomenological analysis
and microscopic multi-agent models
\cite{Mantegna-Stanley-1995-Nature,Plerou-Gopikrishnan-Amaral-Meyer-Stanley-1999-PRE,Plerou-Gopikrishnan-Amaral-Gabaix-Stanley-2000-PRE,Giardina-Bouchaud-Mezard-2001-PA,Ren-Zheng-2003-PLA,Challet-Zhang-1997-PA,Stauffer-Sornette-1999-PA,Muzy-Delour-Bacry-2000-EPJB,Eguiluz-Zimmermann-2000-PRL,Krawiecki-Holyst-Helbing-2002-PRL,Zheng-Qiu-Ren-2004-PRE,Ren-Zheng-Qiu-Trimper-2006-PRE,Gu-Zhou-2009-EPL}.

Recent empirical studies show that the trading volume is highly
correlated with the price return and volatility
\cite{Gabaix-Gopikrishnan-Plerou-Stanley-2003-Nature,Hasbrouck-1991-JF,Karpoff-1987-JFQA,Gopikrishnan-Plerou-Gabaix-Stanley-2000-PRE},
and this confirms the famous saying that it takes trading volume to
move stock prices. A positive linear correlation is revealed based
on the data analysis at time scales large than one minute
\cite{Wood-McInish-Ord-1985-JF,Gallant-Rossi-Tauchen-1992-RFS,Saatcioglu-Starks-1998-IJF}.
For the high-frequency data at microscopic transaction level, the
volume-return relation follows a scaling behavior. Lillo et al.
found a mater curve with scaling form using the Trade and Quote
database of US stocks \cite{Lillo-Farmer-Mantegna-2003-Nature}, and
the scaling function is found to be a power-law form for large
volumes. Lim and Zhou found similar scaling behavior in Australian
and Chinese stocks \cite{Lim-Coggins-2005-QF,Zhou-2007-XXX}. Due to
the significant importance of the trading volume, its statistical
properties is worthy of being carefully analyzed.

Interestingly the cumulative probability distributions of the
trading volume as well as the number of trades also show power-law
behaviors, and the corresponding exponents are reported to be
$\xi_V=1.5$ and $\xi_N=3.4$ respectively
\cite{Plerou-Gopikrishnan-Amaral-Gabaix-Stanley-2000-PRE,Gabaix-Gopikrishnan-Plerou-Stanley-2003-Nature,Gopikrishnan-Plerou-Gabaix-Stanley-2000-PRE}.
To understand the power-law distributions of the trading volume and
number of trades, and their possible relations with the distribution
of price returns, an effective theory is promoted by Gabaix et al
\cite{Gabaix-Gopikrishnan-Plerou-Stanley-2003-Nature}, based on the
empirical power-law relation between the large trading volume and
the large price return. Instead, from the phenomenological analysis
of the order book data, Farmer and Weber et al. suggest that large
price movements are driven by the fluctuation in liquidity
\cite{Farmer-Gillemot-Lillo-Mike-Sen-2004-QF,Weber-Rosenow-2006-QF},
e.g., variations in the response to changes in supply and demand,
and the low density of limit orders, etc. In fact, it is highly
nontrivial to offer a complete answer, how large price movements
occur. Nevertheless, it remains very important to fully understand
the statistical properties of the financial fluctuations, such as
the power-law distributions of different quantities, and their
relations, as well as the long-range time correlations.

In this paper we develop a multi-agent model of trading activity,
aiming at a full understanding of the statistical properties of the
financial fluctuations including the power-law distributions of the
price return, trading volume and number of trades and the long-range
time correlation of the volatility. In the literature, some
stochastic models of trading activity, typically at the
phenomenological level, have been analyzed for this purpose
\cite{Gontis-Kaulakys-2004-PA,Sato-2004-PRE,Das-2005-QF}. Certain
aspects of the financial fluctuations could be reproduced. A
multiplicative stochastic model of the time interval between two
successive trades, for example, is able to reproduce the statistical
properties of the number of trades \cite{Gontis-Kaulakys-2004-PA},
but does not refer to the relation with the power-law distributions
of the price return and trading volume. In the present paper, we
construct our model based on the microscopic structure and
interactions, to capture fundamental mechanisms in the financial
dynamics.

The paper is organized as follows. In Sec. II, the dynamic herding
model with interactions of trading volumes is introduced. In Sec.
III, numerical results of the model are presented. Sec. IV contains
the conclusion.

\section{Model definition}
\label{sec:model} The concept of percolating or herding is important
in describing the financial markets
\cite{Stauffer-Sornette-1999-PA,Eguiluz-Zimmermann-2000-PRL,Cont-Bouchaud-2000-MeD}.
The dynamic version of the static percolation model, the so-called
EZ herding model, shows certain attractive features
\cite{Eguiluz-Zimmermann-2000-PRL}, e.g., the herding structure is
dynamically generated in a simple but robust way. The EZ herding
model captures the power-law distribution of the price return, but
the volatility is short-range correlated in time. To achieve the
long-range time correlation of the volatility, a feed-back
interaction should be introduced
\cite{Zheng-Qiu-Ren-2004-PRE,Zheng-Ren-Trimper-Zheng-2004-PA}. Up to
now, however, it is still far from realistic and harmonic. For
example, only the volatility is concerned. The trading volume and
number of trades have not been touched. The price return is
calculated from the volatility with random $\pm 1$ signs, and this
should not describe the realistic financial dynamics.

In this section, we develop a dynamic herding model including the
price return, trading volume, and number of trades. In fact, it is
not easy to build such a model. We have probed many possible
variations of the microscopic structure and interaction, and finally
come to the present form.

\subsection{Standard EZ herding model}
\label{sec:standard}

To start, let us first consider the EZ herding model. The system
consists of $M$ agents, which form clusters during dynamic
evolution. Initially, each agent is a cluster. The dynamics evolves
in the following way:

(1) At a time step $t$, select an agent $i$ (and thus its cluster)
    at random.

(2) With a probability $1-a$, $i$ remains inactive in trading, and
    select another agent $j$ randomly. If $i$ and $j$ are in different
    clusters, combine the two clusters into one.

(3) With a probability $a$, $i$ becomes active and makes a trade.
    Then all agents in the cluster follow. After that, this cluster
    is broken into a state that each agent is a separate cluster.
    The size of this cluster is recorded as $s(t)$.

Here the probability $a$ is a constant, and controls the dynamic
evolution. Since one does not define buying or selling of the trade,
only the magnitude of the price return defined as $|r(t)|=s(t)$ is
essentially generated. The step (2) represents transmission of
information. Considering the time between two actions as the time
unit, $1/a$ is the rate of transmission of information. If $a$ is
small, for example, transmission of information is fast, and agents
tend to form larger clusters and act collectively. Numerical
simulations \cite{Eguiluz-Zimmermann-2000-PRL} show that for a
certain value of $a$, the probability distribution $P(s)$ obeys a
power law.

\subsection{Herding model interacted with trading volume}
\label{sec:interacted}

However, the EZ herding model does not exhibit the long-range time
correlation of the volatility. Furthermore, we need to include the
price return, trading volume and number of trades
\cite{Gabaix-Gopikrishnan-Plerou-Stanley-2003-Nature,Hasbrouck-1991-JF,Gopikrishnan-Plerou-Gabaix-Stanley-2000-PRE}.
Therefore, we assume that each agent trades with an individual
trading volume $v_{i}(t)> 0$. Denoting buying and selling with
$\sigma_{i}(t)=+1$ and $-1$ respectively, all agents in a cluster
are given a same trade sign $\sigma_{i}(t)$. Initially, each agent
$i$ with $v_{i}(0)=1$ is a cluster, and randomly selects a trade
sign $\sigma_{i}(0)=\pm 1$. The total trading volume is set to
$V(0)=1$. We construct the dynamics as following:

(1) At a time step $t$, select an agent $i$ (and thus its cluster)
    at random, and calculate the trading probability
\begin{equation}
a_{i}(t)=\frac{1}{1+bV(t-1)/v_{i}(t-t')},\quad t'\geq 1 .
\label{e10}
\end{equation}
Here $v_{i}(t-t')$ is the trading volume of $i$  at its last trade,
$V(t-1)$ is the total trading volume at $t-1$, and the parameter $b$
is a positive value.

(2) With a probability $1-a_{i}(t)$, the agent $i$ remains inactive
in trading, and select another agent $j$ randomly.
    If $i$ and $j$ are in different clusters, combine the two clusters into
one. The trade sign of the new cluster is taken to be that of the
larger cluster of the previous two.

(3) With a probability $a_{i}(t)$, all agents in the cluster which
$i$ belongs to and another randomly selected cluster become active,
and make trades according to their trade signs. After that, these
two clusters are broken into a state that each agent is a separate
cluster with a trade sign selected randomly.

Our plausible observation is that if an agent is collecting much
information, i.e., with a small $a_{i}(t)$, it may perform a large
trade. Therefore, we assume $v_{i}(t)=1/a_{i}(t)$ to be the trading
volume of the agent $i$. Then $V(t)=\sum_{i} v_{i}(t)$ is the total
trading volume, with the sum over the two active clusters. Let
$N(t)$ denote the number of agents in the two active clusters, we
define it as the number of trades.

To determine the price return, we need more careful consideration.
Empirical studies show that the trading volume seems to have a
square root impact on the price return, and the price return
saturates at extremely large trading volumes
\cite{Gabaix-Gopikrishnan-Plerou-Stanley-2003-Nature,Hasbrouck-1991-JF}.
Further, the correlation between the price return and trading volume
is largely due to the number of trades
\cite{Gopikrishnan-Plerou-Gabaix-Stanley-2000-PRE,Plerou-Gopikrishnan-Gabaix-Stanley-2002-PRE}.Following
the square root price impact function, we assume that the price
return is determined by the volume imbalance $Q(t)=\sum_{i}
v_{i}(t)\sigma_{i}$ and the number of trades. The volume imbalance
reflects the difference between supply and demand
\cite{Farmer-Gillemot-Lillo-Mike-Sen-2004-QF,Weber-Rosenow-2006-QF}.
Quantitatively, we define the price return
\begin{equation}
r(t)=Sign(Q(t))\frac{\sqrt{|Q(t)|}}{\sqrt{|Q(t)|}+A} \sqrt{N(t)}.
\label{e20}
\end{equation}
The parameter $A$ is taken to be a large positive value, such that
$r(t)\sim\sqrt{|Q(t)|}\sqrt{N(t)}$ at relatively small $|Q(t)|$, and
$r(t)\sim \sqrt{N(t)}$ at extremely large $|Q(t)|$.

The key ingredient in our model is the time-dependent probability
$a_i(t)$. In Eq.~({\ref {e10}}), we assume that $a_{i}(t)$ depends
on the ratio of the total trading volume at time $t-1$ to its
individual trading volume at its last trade. In financial markets, a
large trading volume is usually accompanied by the strong
fluctuation of the price return
\cite{Gabaix-Gopikrishnan-Plerou-Stanley-2003-Nature,Hasbrouck-1991-JF,Gopikrishnan-Plerou-Gabaix-Stanley-2000-PRE}.
This inversely leads to large trading volumes in next time steps.
Therefore, $a_{i}(t)$ is taken to be inversely proportional to the
trading volume $V(t-1)$. If $V(t-1)$ is large, transmission of
information is fast, the probability of combining two clusters is
high, then the number of the trades increases on average, and
finally leads to large trading volumes. Such a dynamic feed-back
interaction of the trading volume essentially generates the
long-range time correlation of the volatility. On the other hand,
$a_{i}(t)$ is taken to be proportional to the individual trading
volume $v_{i}(t-t')$ at its last trade, based on the empirical
assumption that an agent with a large trading volume in its last
trade may be more active in trading in next time steps. Further, the
thermodynamic limit is well defined due to the ratio
$V(t-1)/v_{i}(t-t')$ in Eq.~({\ref {e10}}).

\section{Simulation results}
\label{sec:results}

\subsection{Probability distribution function}
\label{sec:PDF}

\begin{figure}[htb]
\centering
\includegraphics[width=7cm]{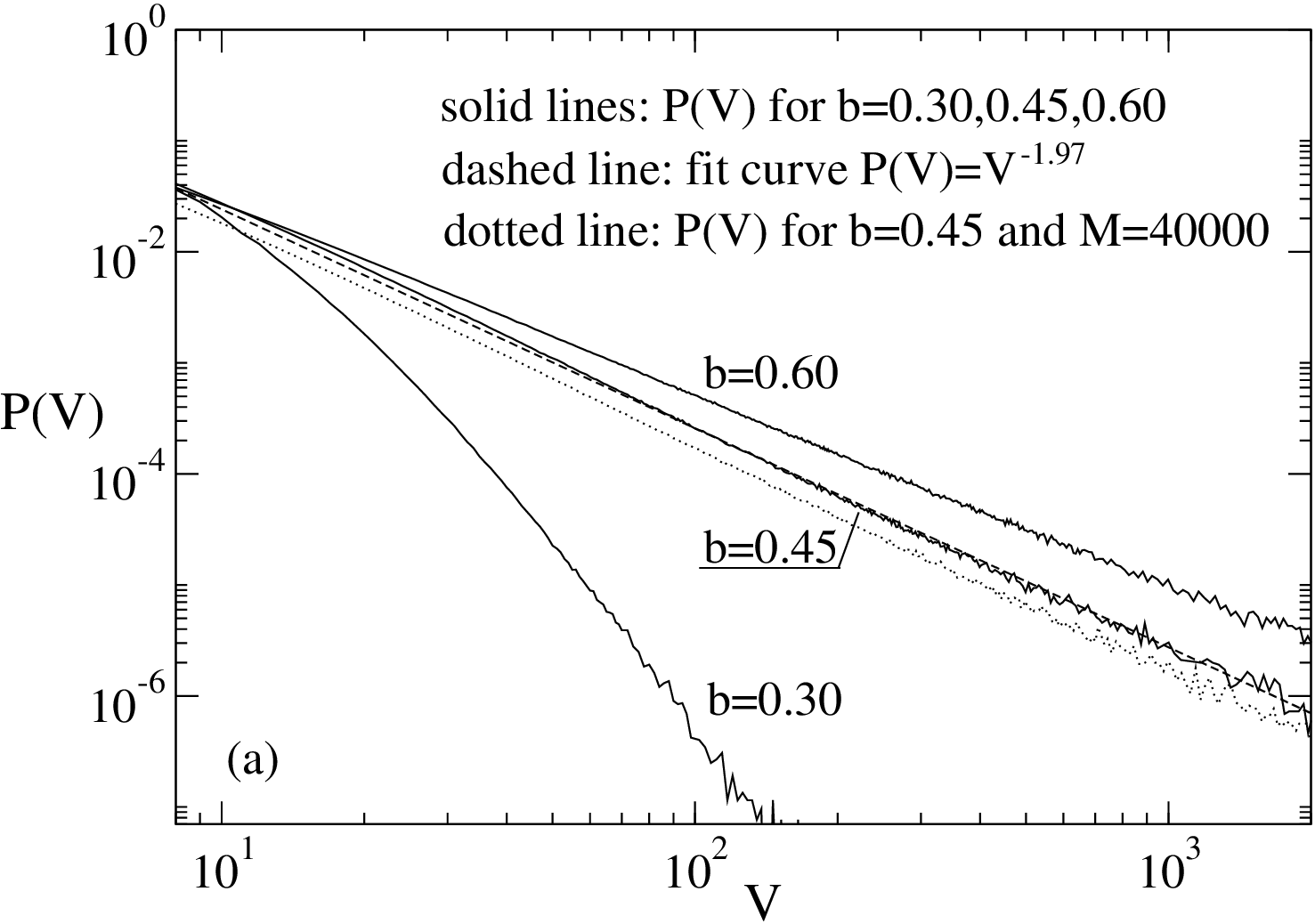}
\includegraphics[width=7cm]{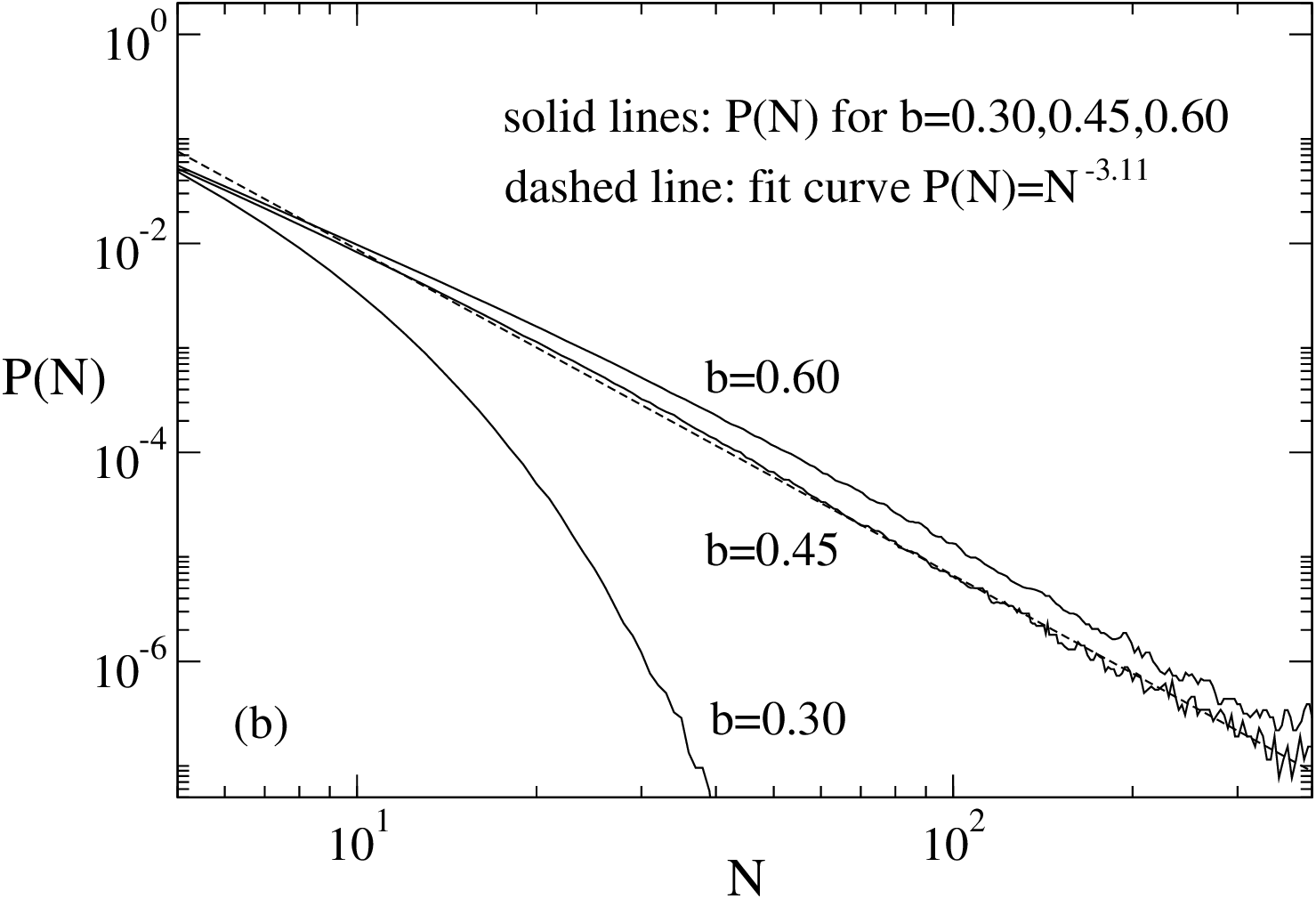}
\caption{(Color online) Probability distributions of the trading
volume and number of trades are plotted for $b=0.30, 0.45$ and
$0.60$ in (a) and (b). The results are obtained with a total number
of agents $M=80000$. For comparison, a curve of $b=0.45$ with
$M=40000$ is also displayed in (a).} \label{Fig:PDF:volum:number}
\end{figure}

In our model, the only tunable parameter is $b$. In calculating the
price return, we fix the value $A=50$. For each value of $b$, we
take an average over $10^{8}$ iterations, after $10^{6}$ iterations
for equilibration. To detect the finite size effect, we perform
extensive simulations with different total numbers of agents, and
find that the results become stable for $M\geq40000$ as shown in
Fig. \ref {Fig:PDF:volum:number} (a). Therefore, we report the
results with $M=80000$.

From empirical studies of the stock time series, the probability
distribution of the trading volume $V$ obeys a power law
\begin{equation}
P(V) \sim V^{-(1+\xi_{V})}, \label{e30}
\end{equation}
with $\xi_{V}=1.5$, while that of the number of trades $N$ obeys
\begin{equation}
P(N) \sim N^{-(1+\xi_{N})}, \label{e40}
\end{equation}
with $\xi_{N}=3.4$. The exponents of these two power-law
distributions appear to have an approximate relation $\xi_{N} \cong
2\xi_{V}$ \cite{Gabaix-Gopikrishnan-Plerou-Stanley-2003-Nature} .

The probability distributions of the trading volume and number of
trades in our model are carefully investigated. In Fig. \ref
{Fig:PDF:volum:number} (a) and (b), $P(V)$ and $P(N)$ are plotted
for $b=0.30,0.45,0.60$ in log-log scale. For a small $b$, e.g.,
$b\leq 0.3$, $P(V)$ and $P(N)$ decay rapidly, and do not show a
power-law behavior. As $b$ increases, both $P(V)$ and $P(N)$ show a
power-law behavior at $b=0.45$, at least in two orders of magnitude.
In this sense, the system exhibits a 'cross-over' behavior. Fitting
the curves with the power laws in Eqs.(\ref{e30}) and (\ref{e40}),
we estimate $\xi_{V}=0.97$ and $\xi_{N}=2.11$. These values of the
exponents are consistent with the approximate relation $\xi_{N}
\cong 2\xi_{V}$. For $b>0.45$, $P(V)$ remains a power-law behavior
in a broad range of $b$, but the exponent $\xi_{V}$ changes with $b$
and becomes smaller. On the other hand, $P(N)$ deviates from a
power-law behavior up to a medium value $N$, while a power-law tail
is still kept with an exponent $\xi_{N}$ approximately the same as
that at $b=0.45$. In Fig. \ref {Fig:PDF:volum:number}, the curves
for $b=0.60$ are displayed.\\

\begin{figure}[htb]
\centering
\includegraphics[width=7cm]{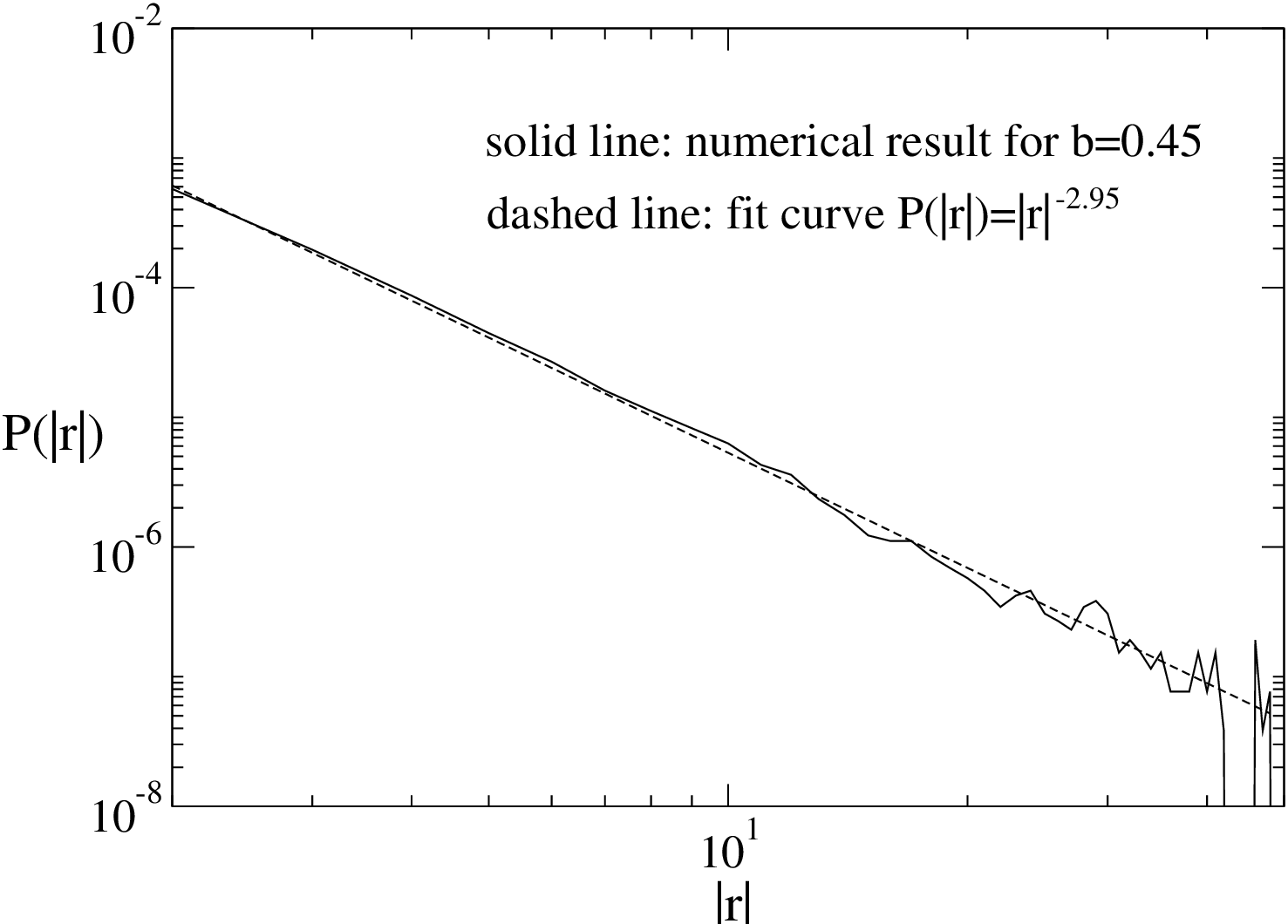}
\caption{(Color online) Probability distribution of the volatility
for $b=0.45$ and $M=80000$.} \label{Fig:PDF:volatility}
\end{figure}

It is well known from the empirical analysis that the probability
distribution of the price return exhibits a power-law tail
\begin{equation}
P(|r|) \sim |r|^{-(1+\xi_{r})}, \label{e50}
\end{equation}
with $\xi_{r}=3.0$
\cite{Plerou-Gopikrishnan-Amaral-Gabaix-Stanley-2000-PRE}. In Fig.
\ref {Fig:PDF:volatility}, the probability distribution of the price
return of our model is plotted for $b=0.45$ in log-log scale. The
curve can be nicely fitted with a power law, and the exponent
$\xi_{r}$ is estimated to be $1.95$, close to $\xi_{N}=2.11$. In
summary, our dynamic herding model at $b=0.45$ captures the
power-law distributions of the trading volume, number of trades and
price return. The corresponding exponents of these power-law
distributions follow an approximate relation $\xi_{r} \cong \xi_{N}
\cong 2\xi_{V}$. This is in agreement with that of the real markets
reported in Ref.
\cite{Gabaix-Gopikrishnan-Plerou-Stanley-2003-Nature}, though the
exponents themselves are slightly different.

The specific form of $a_{i}(t)$ is not very important. Other
functions may also work, if their behaviors are similar to that in
Eq.~(\ref{e10}). To verify this we also study the model with another
$a_{i}(t)$ with a form
\begin{equation}
a_{i}(t)=1-ce^{-d\frac{V(t-1)}{v_{i}(t-t')}},\quad t'\geq 1 ,
\label{e60}
\end{equation}
where $c$ and $d$ are two positive parameters. We remain the inverse
relation between $a_{i}(t)$ and the ratio $V(t-1)/v_{i}(t-t')$, thus
makes it behave similar to Eq.~({\ref {e10}}). An important
characteristic of this model is the simple relation
$v_{i}(t)=1/a_{i}(t)$ between the trading volume and trading
probability. It indicates that $v_{i}(t)$ are the dynamic variables,
interact each other through the trading probability $a_{i}(t)$, and
evolve according to Eq.~(\ref{e60}).

\begin{figure}[htb]
\centering
\includegraphics[width=7cm]{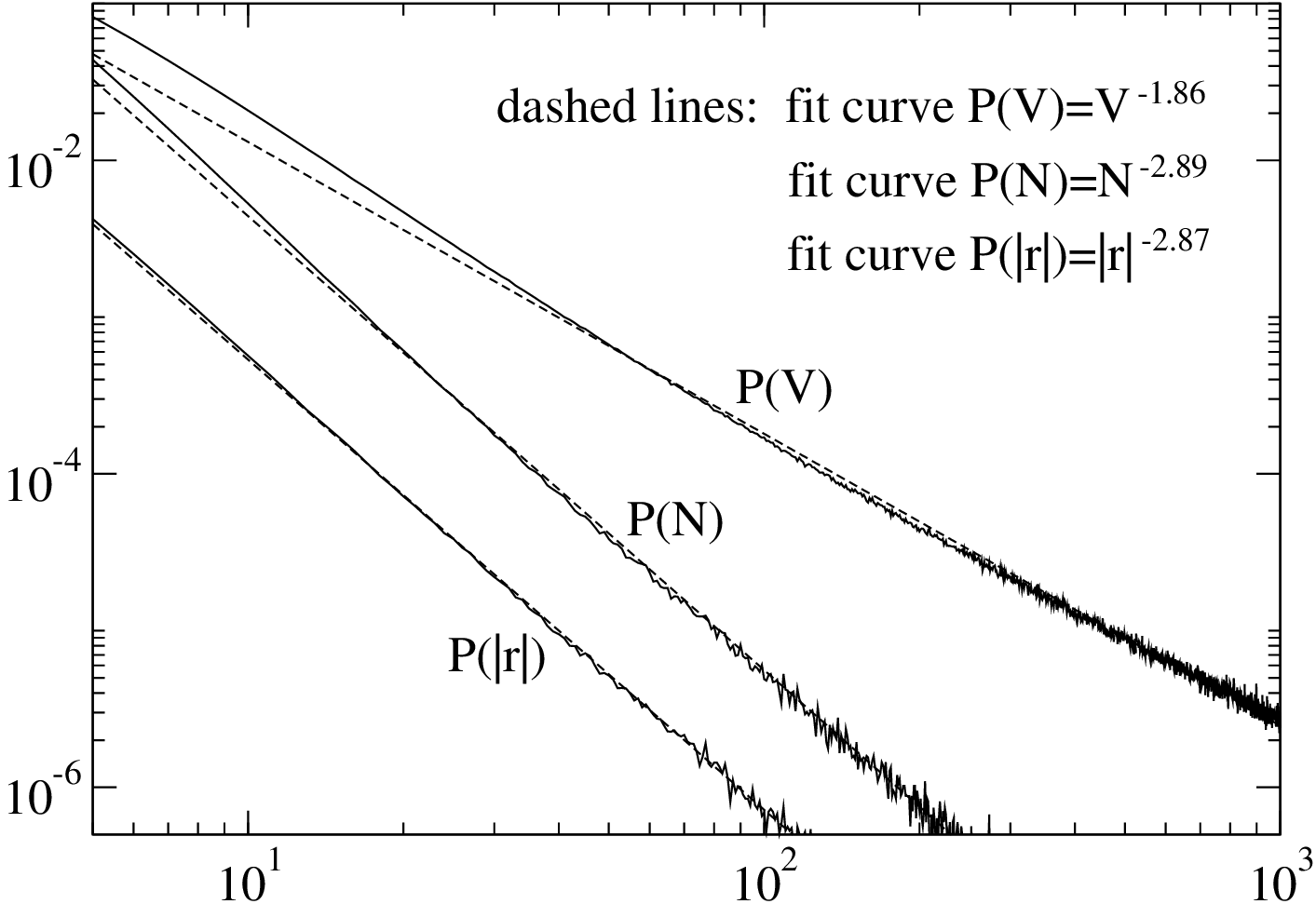}
\caption{(Color online) Probability distributions of trading volume,
number of trades, and price return for the model with $a_{i}(t)$
defined in Eq.~(\ref{e60}). Results are obtained with $c=1.0$,
$d=2.0$ and $M=80000$.} \label{Fig:PDF:a1}
\end{figure}

We study the probability distributions of trading volume, number of
trades and price return for the model with $a_{i}(t)$ defined in
Eq.~(\ref{e60}). To make the probability $a_{i}(t)$ fixed between
$0$ and $1$, we set $c=1.0$. By adjusting the parameter $d$, one
observes a 'cross-over' behavior similar to the model with
$a_{i}(t)$ defined in Eq.~({\ref {e10}}): for small $c$, no
power-law behavior is observed, while for large $c$ power-law
behavior occurs. In Fig. \ref{Fig:PDF:a1}, power behaviors of
$P(V)$, $P(N)$ and $P(|r|)$ for a large $c=2.0$ are plotted. By
fitting the slopes of these curves, we estimate $\xi_{V}=0.86$,
$\xi_{N}=1.89$ and $\xi_{r}=1.87$. These exponents display a
relation $\xi_{r} \cong \xi_{N} \cong 2\xi_{V}$ also consistent with
that of the real markets.\\\\

\subsection{Auto-correlation function}
\label{sec:correlation}

The long-range time correlation of the volatility is another
important feature of the financial markets. Let us define the the
auto-correlation function of the magnitude of the price return as
\begin{equation}
C(\tau)\equiv \frac{<|r(t)||r(t+\tau)|> - (<|r(t)|>)^2}{<|r(t)|^2> -
(<|r(t)|>)^2 }, \label{e70}
\end{equation}
where $<\cdot\cdot\cdot>$ represents the average over the time $t$.
From the empirical analysis, it obeys a power law,
\begin{equation}
C(\tau)\sim \tau^{-\lambda}, \label{e80}
\end{equation}
with $\lambda\cong0.3$
\cite{Liu-Gopikrishnan-Cizeau-Meyer-Peng-Stanley-1999-PRE,Giardina-Bouchaud-Mezard-2001-PA}.

\begin{figure}[htb]
\centering
\includegraphics[width=7cm]{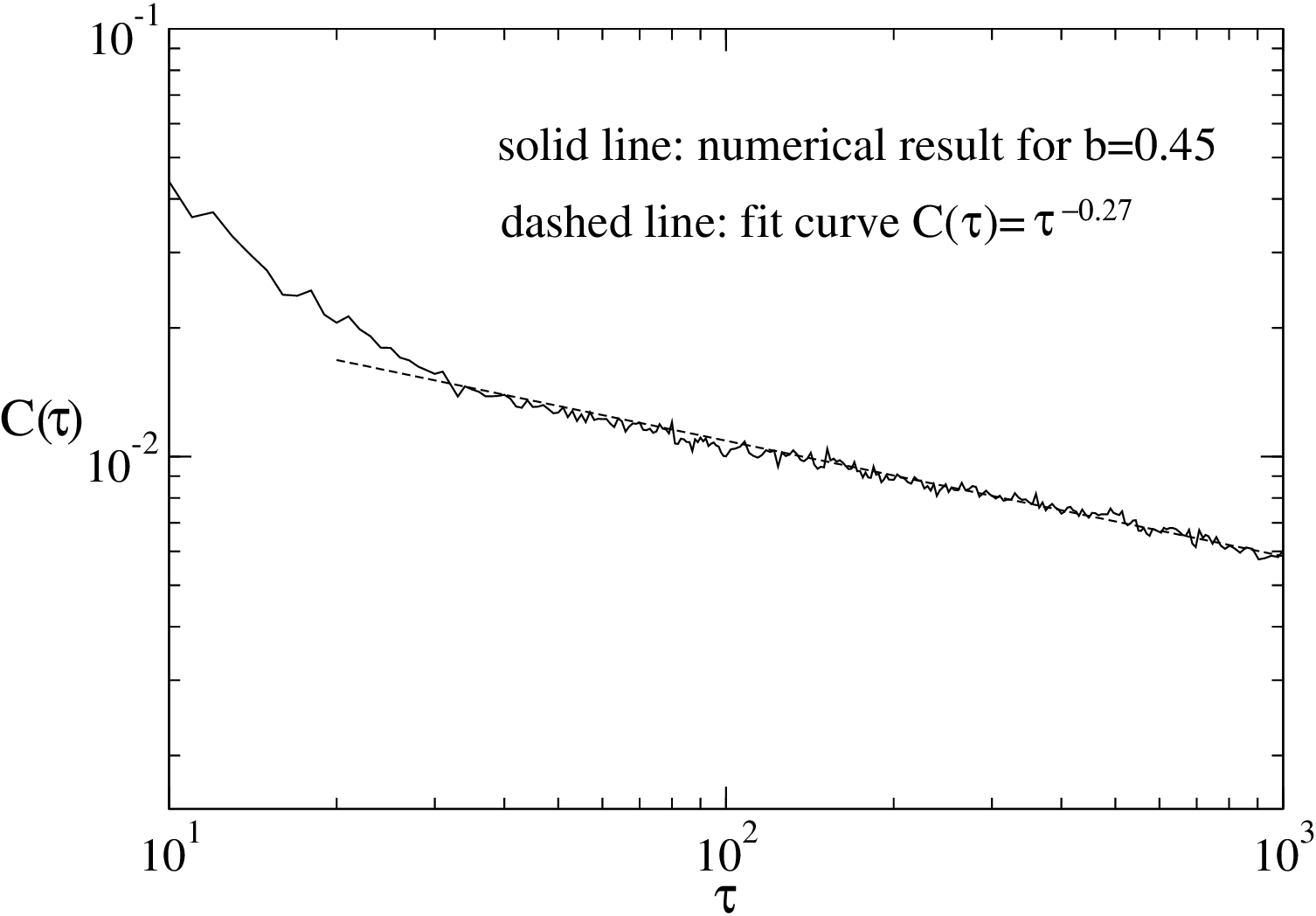}
\caption{(Color online) auto-correlation function of the volatility
for $b=0.45$ and $M=80000$.} \label{Fig:auto:volatility}
\end{figure}

We calculate $C(\tau)$ of the model with $a_{i}(t)$ defined in
Eq.~(\ref{e10}), and observe that a power-law behavior is achieved
only when $b$ is in the very neighborhood of $0.45$. In Fig. \ref
{Fig:auto:volatility} (b), $C(\tau)$ is plotted for $b=0.45$ in
log-log scale. The curve can be nicely fitted by a power law,
indicating a long-range time correlation of the volatility. The
exponent $\lambda$ is estimated to be $0.27$, very close to that of
the real markets. This improves the result $\lambda=0.90$ in a naive
model \cite{Zheng-Ren-Trimper-Zheng-2004-PA}. A power law with an
exponent $0.53$ (not shown in figure) is also observed in $C(\tau)$
of the model with another $a_{i}(t)$ defined in Eq.~(\ref{e60}), but
with a reactively bigger fluctuation.\\

In our model, a large price movement is induced by a large volume
imbalance $|Q(t)|$ and a large number of trades $N(t)$ due to the
price dynamics defined in Eq.~(\ref{e20}). This may occur when large
clusters exist in the system. Large clusters are formed after agents
are active in collecting information for a period of time. However,
a single large trading volume does not necessarily lead to a large
price movement.

\section{Conclusion}
\label{sec:concl}

We introduce a dynamic herding model with interactions of trading
volumes. At time $t$, each agent trades with a probability
$a_{i}(t)$ as a linear function of the ratio of the total trading
volume $V(t-1)$ at time $t-1$ to its own trading volume
$v_{i}(t-t')$ at its last trade. Agents are endowed with trade signs
$\sigma_i(t)=\pm 1$ denoting buying and selling, and the price
return is determined by the volume imbalance and number of trades.
We find that at a threshold $b=0.45$, our model reproduces the
power-law distributions of the trading volume, number of trades and
price return, and more importantly, the approximate relation
$\xi_{r} \cong \xi_{N} \cong 2\xi_{V}$. The generated volatilities
are long-range correlated in time. We also investigate the model
with another exponential form of $a_{i}(t)$ which remains its
inverse relation with the ratio ${V(t-1)}/{v_{i}(t-t')}$, and
similar power-law behaviors and long-range correlation are observed.
This indicates the robustness of our model.

By simulating the agents' reactions to the market information
through the interactions between their individual trading volumes
and the total trading volumes, the model can explain the power-law
behaviors of the trading volume, number of trades and price return,
and their relations. We believe that our model captures certain
essences of the financial markets. Further works should include
understanding of the Leverage and anti-Leverage effects in western
and Chinese financial markets
\cite{Ren-Zheng-Lin-Wen-Trimper-2005-PA,Qiu-Zheng-Ren-Trimper-2006-PRE,Qiu-Zheng-Ren-Trimper-2007-PA,Shen-Zheng-2009B-EPL},
and more general interactions between the price return, trading
volume and number of trades, etc.

\bigskip
{\textbf{Acknowledgments:}}

This work was supported in part by NNSF (China) under grant Nos. 10325520,
70371069 and 10905023, and "Chen Guang" project sponsored by
Shanghai Municipal Education Commission and Shanghai Education
Development Foundation under grant No. 2008CG37.

\bibliography{E:/Papers/Auxiliary/Bibliography}

\begin{thebibliography}{10}
\expandafter\ifx\csname url\endcsname\relax
  \def\url#1{\texttt{#1}}\fi
\expandafter\ifx\csname urlprefix\endcsname\relax\def\urlprefix{URL }\fi
\expandafter\ifx\csname href\endcsname\relax
  \def\href#1#2{#2} \def\path#1{#1}\fi

\bibitem{Mandelbrot-1963-JB}
B.~B. Mandelbrot, The variation of certain speculative prices, J. Business 36
  (1963) 394--419.

\bibitem{Fama-1965-JB}
E.~F. Fama, The behavior of stock market prices, J. Business 38~(1) (1965)
  34--105.

\bibitem{Mantegna-Stanley-1995-Nature}
R.~N. Mantegna, H.~E. Stanley, {Scaling behaviour in the dynamics of an
  economic index}, Nature 376 (1995) 46--49.

\bibitem{Lux-1996-AFE}
T.~Lux, {The Stable Paretian Hypothesis and the Frequency of Large Returns: An
  Examination of Major German Stocks}, Appl. Financ. Econ. 6 (1996) 463--475.

\bibitem{Ghashghaie-Breymann-Peinke-Talkner-Dodge-1996-Nature}
S.~Ghashghaie, W.~Breymann, J.~Peinke, P.~Talkner, Y.~Dodge, {Turbulent
  cascades in foreign exchange markets}, Nature 381 (1996) 767--770.

\bibitem{Plerou-Gopikrishnan-Amaral-Meyer-Stanley-1999-PRE}
V.~Plerou, P.~Gopikrishnan, L.~A.~N. Amaral, M.~Meyer, H.~E. Stanley, {Scaling
  of the distribution of price fluctuations of individual companies}, Phys.
  Rev. E 60 (1999) 6519--6529.

\bibitem{Plerou-Gopikrishnan-Amaral-Gabaix-Stanley-2000-PRE}
V.~Plerou, P.~Gopikrishnan, L.~A.~N. Amaral, X.~Gabaix, H.~E. Stanley,
  {Economic fluctuations and anomalous diffusion}, Phys. Rev. E 62 (2000)
  R3023--R3026.

\bibitem{Gabaix-Gopikrishnan-Plerou-Stanley-2003-Nature}
X.~Gabaix, P.~Gopikrishnan, V.~Plerou, H.~E. Stanley, {A theory of power-law
  distributions in financial market fluctuations}, Nature 423 (2003) 267--270.

\bibitem{Giardina-Bouchaud-Mezard-2001-PA}
I.~Giardina, J.~Bouchaud, M.~M{\'e}zard, {Microscopic models for long ranged
  volatility correlations}, Physica A 299 (2001) 28--39.

\bibitem{Liu-Gopikrishnan-Cizeau-Meyer-Peng-Stanley-1999-PRE}
Y.-H. Liu, P.~Gopikrishnan, P.~Cizeau, M.~Meyer, C.-K. Peng, H.~E. Stanley,
  {Statistical properties of the volatility of price fluctuations}, Phys. Rev.
  E 60 (1999) 1390--1400.

\bibitem{Mantegna-Stanley-2000}
R.~N. Mantegna, H.~E. Stanley, {An Introduction to Econophysics: Correlations
  and Complexity in Finance}, Cambridge University Press, Cambridge, 2000.

\bibitem{Bouchaud-Potters-2000}
J.-P. Bouchaud, M.~Potters, {Theory of Financial Risks: From Statistical
  Physics to Risk Management}, Cambridge University Press, Cambridge, 2000.

\bibitem{Sornette-2003}
D.~Sornette, {Why Stock Markets Crash: Critical Events in Complex Financial
  Systems}, Princeton University Press, Princeton, 2003.

\bibitem{Ren-Zheng-2003-PLA}
F.~Ren, B.~Zheng, {Generalized persistence probability in a dynamic economic
  index}, Phys. Lett. A 313 (2003) 312--315.

\bibitem{Challet-Zhang-1997-PA}
D.~Challet, Y.-C. Zhang, {Emergence of cooperation and organization in an
  evolutionary game}, Physica A 246 (1997) 407--418.

\bibitem{Stauffer-Sornette-1999-PA}
D.~Stauffer, D.~Sornette, {Self-organized percolation model for stock market
  fluctuations}, Physica A 271 (1999) 496--506.

\bibitem{Muzy-Delour-Bacry-2000-EPJB}
J.-F. Muzy, J.~Delour, E.~Bacry, {Modelling fluctuations of financial time
  series: from cascade process to stochastic volatility model }, Eur. Phys. J.
  B 17 (2000) 537--548.

\bibitem{Eguiluz-Zimmermann-2000-PRL}
V.~M. Egu{\'i}luz, M.~G. Zimmermann, {Transmission of information and herd
  behavior: An application to financial markets}, Phys. Rev. Lett. 85 (2000)
  5659--5662.

\bibitem{Krawiecki-Holyst-Helbing-2002-PRL}
A.~Krawiecki, J.~A. Holyst, D.~Helbing, {Volatility clustering and scaling for
  financial time series due to attractor bubbling}, Phys. Rev. Lett. 89 (2002)
  158701.

\bibitem{Zheng-Qiu-Ren-2004-PRE}
B.~Zheng, T.~Qiu, F.~Ren, {Two-phase phenomena, minority games, and herding
  models}, Phys. Rev. E 69 (2004) 046115.

\bibitem{Ren-Zheng-Qiu-Trimper-2006-PRE}
F.~Ren, B.~Zheng, T.~Qiu, S.~Trimper, {Minority games with score-dependent and
  agent-dependent payoffs}, Phys. Rev. E 74 (2006) 041111.

\bibitem{Gu-Zhou-2009-EPL}
G.-F. Gu, W.-X. Zhou, {Emergence of long memory in stock volatilities from a
  modified Mike-Farmer model}, EPL 86 (2009) 48002.
\newblock \href {http://dx.doi.org/10.1209/0295-5075/86/48002}
  {\path{doi:10.1209/0295-5075/86/48002}}.

\bibitem{Hasbrouck-1991-JF}
J.~Hasbrouck, {Measuring the information content of stock trades}, J. Financ.
  46 (1991) 179--207.

\bibitem{Karpoff-1987-JFQA}
J.~M. Karpoff, {The relation between price changes and trading volume: A
  survey}, J. Financ. Quant. Anal. 22 (1987) 109--126.

\bibitem{Gopikrishnan-Plerou-Gabaix-Stanley-2000-PRE}
P.~Gopikrishnan, V.~Plerou, X.~Gabaix, H.~E. Stanley, {Statistical properties
  of share volume traded in financial markets}, Phys. Rev. E 62 (2000)
  R4493--R4496.

\bibitem{Wood-McInish-Ord-1985-JF}
R.~A. Wood, T.~H. McInish, J.~K. Ord, {An investigation of transactions data
  for NYSE stocks}, J. Financ. 40 (1985) 723--739.

\bibitem{Gallant-Rossi-Tauchen-1992-RFS}
A.~R. Gallant, P.~E. Rossi, G.~Tauchen, {Stock prices and volume}, Rev. Financ.
  Stud. 5 (1992) 199--242.

\bibitem{Saatcioglu-Starks-1998-IJF}
K.~Saatcioglu, L.~T. Starks, {The stock price-volume relationship in emerging
  stock markets: the case of Latin America}, Int. J. Forecast. 14 (1998)
  215--225.

\bibitem{Lillo-Farmer-Mantegna-2003-Nature}
F.~Lillo, J.~D. Farmer, R.~Mantegna, {Master curve for price impact function},
  Nature 421 (2003) 129--130.

\bibitem{Lim-Coggins-2005-QF}
M.~Lim, R.~Coggins, {The immediate price impact of trades on the Australian
  Stock Exchange}, Quant. Financ. 5 (2005) 365--377.

\bibitem{Zhou-2007-XXX}
W.-X. Zhou, {Universal price impact functions of individual trades in an
  order-driven market}, arXiv: 0708.3198v2 (2007).

\bibitem{Farmer-Gillemot-Lillo-Mike-Sen-2004-QF}
J.~D. Farmer, L.~Gillemot, F.~Lillo, S.~Mike, A.~Sen, {What really causes large
  price changes?}, Quant. Financ. 4 (2004) 383--397.

\bibitem{Weber-Rosenow-2006-QF}
P.~Weber, B.~Rosenow, {Large stock price changes: Volume or liquidity?}, Quant.
  Financ. 6 (2006) 7--14.

\bibitem{Gontis-Kaulakys-2004-PA}
Gontis-Kaulakys-2004-PA, {Multiplicative point process as a model of trading
  activity}, Physica A 343 (2004) 505--514.

\bibitem{Sato-2004-PRE}
A.~H. Sato, {Explanation of power law behavior of autoregressive conditional
  duration processes based on the random multiplicative process}, Phys. Rev. E
  69 (2004) 047101.

\bibitem{Das-2005-QF}
S.~Das, {A Learning Market-Maker in the Glosten-Milgrom Model}, Quant. Financ.
  5 (2005) 169.

\bibitem{Cont-Bouchaud-2000-MeD}
R.~Cont, J.-P. Bouchaud, {Herd behavior and aggregate fluctuations in financial
  markets}, Macroecon. Dyn. 4 (2000) 170--196.

\bibitem{Zheng-Ren-Trimper-Zheng-2004-PA}
B.~Zheng, F.~Ren, S.~Trimper, D.-F. Zheng, {A generalized dynamic herding model
  with feed-back interactions}, Physica A 343 (2004) 653--661.

\bibitem{Plerou-Gopikrishnan-Gabaix-Stanley-2002-PRE}
V.~Plerou, P.~Gopikrishnan, X.~Gabaix, H.~E. Stanley, {Quantifying stock-price
  response to demand fluctuations}, Phys. Rev. E 66 (2002) 027104.
\newblock \href {http://dx.doi.org/10.1103/PhysRevE.66.027104}
  {\path{doi:10.1103/PhysRevE.66.027104}}.

\bibitem{Ren-Zheng-Lin-Wen-Trimper-2005-PA}
F.~Ren, B.~Zheng, H.~Lin, L.~Wen, S.~Trimper, {Persistence probabilities of the
  German DAX and Shanghai index}, Physica A 350 (2005) 439.

\bibitem{Qiu-Zheng-Ren-Trimper-2006-PRE}
T.~Qiu, B.~Zheng, F.~Ren, S.~Trimper, {Return-volatility correlation in
  financial dynamics}, Phys. Rev. E 73 (2006) 065103(R).

\bibitem{Qiu-Zheng-Ren-Trimper-2007-PA}
T.~Qiu, B.~Zheng, F.~Ren, S.~Trimper, {Statistical properties of German Dax and
  Chinese indices}, Physica A 378 (2007) 387--398.

\bibitem{Shen-Zheng-2009B-EPL}
J.~Shen, B.~Zheng, {On return-volatility correlation in financial dynamics},
  Europhys. Lett. 88 (2009) 28003.

\end{thebibliography}

\end{document}